\documentclass[pre,amsmath,amssymb]{revtex4}
\usepackage{graphicx}%
\usepackage{amsmath}
\usepackage{amssymb}

\begin{document}

\title{Propagation of Bessel beams from a dielectric to a conducting medium}
\author{D. Mugnai}
\affiliation{ ``Nello Carrara" Institute of Applied Physics,  CNR Florence Research Area \\
 Via Madonna del Piano 10, 
50019 Sesto Fiorentino, Italy}

\begin{abstract}
\vspace{.5cm}
 Recently, the use of Bessel beams in evaluating  the
possibility of using them for a new generation of GPR (ground
penetrating radar) systems has been considered. Therefore, an
analysis of the propagation of Bessel beam in conducting media is
worthwhile. We present here an analysis of this type. Specifically,
for normal incidence we analyze the propagation of a Bessel beam
coming from a perfect dielectric and impinging on a conducting
medium, i.e. the propagation of a Bessel beam generated by refracted
inhomogeneous waves. The remarkable and unexpected result is that
the incident Bessel beam does not change its shape even when
propagating in the conducting medium.
\end{abstract}

 \maketitle

In the last decades, electromagnetic waves have found many
practical applications, including ones that are different from
their traditional utilization in the field of communications.
Among these applications, the detection and characterization of
buried objects of a non-metallic nature is of particular interest.
In relation to this topic, Bessel beams have recently been
considered for evaluating the possibility of using  localized
waves in GPR systems.

Bessel beams (also known as localized Bessel-X waves) are of great
interest in physics because of their characteristics of being non
diffracting beams, and due to their implications with regard to the
topic of superluminality \cite{all}.

Apart from these features, their nature of localized waves may make
them suitable also for other field of interest, As mentioned
previously, the possibility of using  Bessel beams in GPR apparatus
has recently been considered \cite{mug1}. This type of utilization
requires a knowledge of the propagation of the beam in conducting,
absorbing, media \cite{zam}.

The aim of this note is to analyze the propagation of a Bessel beam
coming from a dielectric medium (air, for example) and impinging at
normal incidence into a conducting medium, i.e. the propagation of a
Bessel beam generated by refracted inhomogeneous waves. We will
demonstrate that, quite surprisingly, the beam does not change its
shape when propagating in the conducting medium, but solely
attenuates by going far away from the interface.

The following analysis  works in the scalar approximation, since
it refers to the specific system with specific sources. For this
system, it has been demonstrated that, among the three components
of the field, there is one which is dominant with respect to the
other two. In this situation, the scalar approximation works well,
and thus its use is justified \cite{mug}.

\begin{figure}[h,t]
\begin{center}
\includegraphics[width=7 cm]{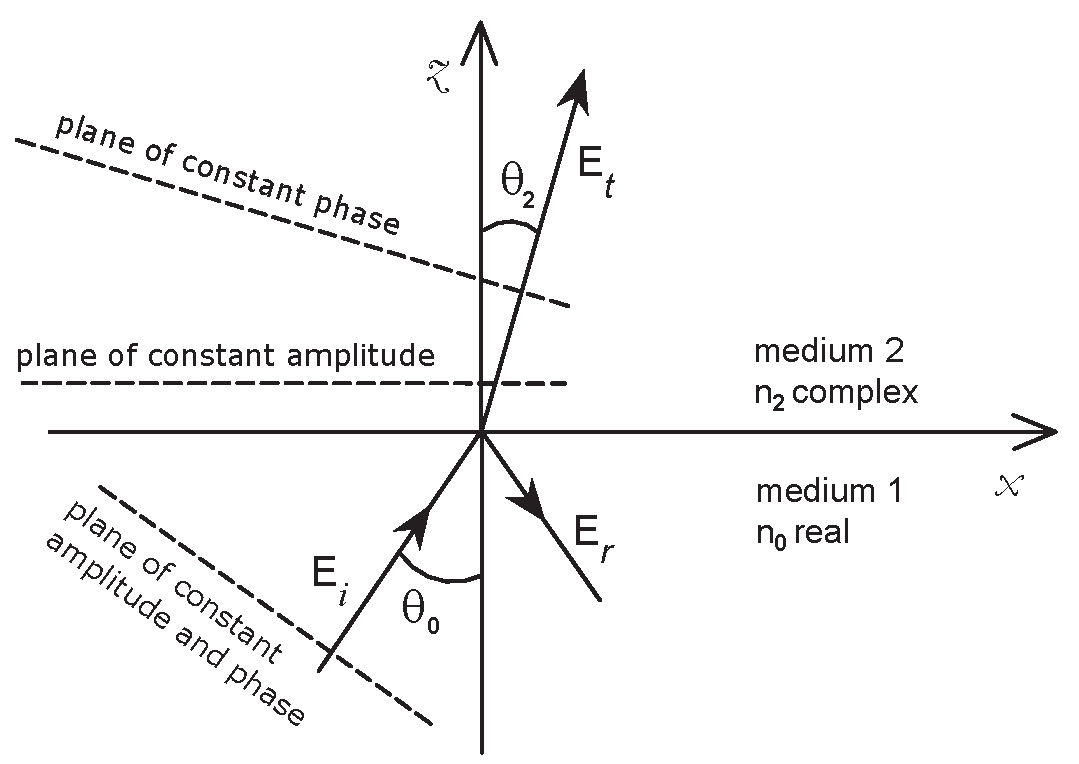}
 \caption{Scheme of the propagation of a plane wave
coming from medium 1 and impinging on medium 2:  medium 1 is  a
perfect dielectric, while medium 2 is a conducting medium. The two
media (which are separated by a plane interface) are supposed to be
homogeneous and of infinite width. $\theta_0$ and $\theta_2$ are the
incident (real) and refracted (complex) angles, respectively.
$E_i,\,E_r$, and $E_t$ denote the incident, reflected and
transmitted fields, respectively.  In the dielectric medium the
planes of constant amplitude and phase are both tilted with respect
to the interface. On the contrary, in the conducting medium only the
planes of constant phase are tilted with respect
 to the interface, while the planes of constant amplitude are parallel to the surface.}
 \label{scheme}
 \end{center}
 \end{figure}

Let us begin by considering a system formed by two half-spaces, 1
and 2,  as sketched in Fig. \ref{scheme}.

A plane wave coming from medium 1 impinges on medium 2 with the
incident angle $\theta_0$. Let medium 1 be a perfect dielectric, and
medium  2 a conducting medium that is separated from medium 1 by a
plane interface. The incident plane wave $E_i$ in medium 1 is

\begin{equation}
\label{ei}
 E_i=\exp{[ik_0(\alpha x+\beta y +\gamma z)]}  .
\end{equation}
This impinging plane wave is expected to give rise in medium 2 to
a plane wave $E_t$, but of  complex type. This is due to the fact
that the refractive index $n_2$ in medium 2 is complex, hence an
exponential decrease of the real amplitude is expected. Thus in
medium 2 we write

\begin{eqnarray}
\label{et}
  E_t = T \exp{[ik_2(\alpha_2 x+\beta_2 y
+\gamma_2  z)]} =  T \exp{[ik_0n_2(\alpha_2 x+\beta_2 y +\gamma_2
z )]},
\end{eqnarray}
where $k_0$ and $k_2\:(=k_0n_2)$ are the wave numbers in medium 1
(vacuum or air) and in medium 2, respectively; $T$ is the
transmission coefficient, which can be found by applying the
boundary conditions at the interface between the two media; $\alpha
,\beta , \gamma $ are the real cosine directors of the direction of
propagation in medium 1, while $\alpha_2 ,\beta_2 , \gamma_2 $ are
constants only subject to the condition that
\begin{equation}
\alpha_2^2+\beta_2^2+\gamma_2^2=1\:.
\end{equation}
For the sake of simplicity, we have taken the amplitude of the
incoming wave as being equal to one, and we have omitted the
temporal factor $\exp(-i\omega t)$. In a system of spherical
coordinates, the quantities  $\alpha ,\beta , \gamma $ and $\alpha_2
,\beta_2 , \gamma_2 $ may be written as

\begin{align}
\alpha =&\sin\theta_0\cos\psi \qquad  &   \qquad    &\alpha_2=\sin\theta_2\cos\psi_2 \nonumber \\
 \beta =&\sin\theta_0\sin\psi         &              &\beta_2=\sin\theta_2\sin\psi_2  \\
 \gamma =&\cos\theta_0                &             &\gamma_2=\cos\theta_2\:,     \nonumber
\end{align}

In principle, $n_2\alpha_2,\:n_2\beta_2$, and $\psi_2$ could be
complex quantities. However, this is not so, as can be inferred by
simple physical considerations.

As is well known \cite{str,bor}, in the propagation through a plane
interface into a conducting medium, the field is described by a
system of inhomogeneous equations. Inside the conducting medium the
planes of constant amplitude are parallel to the interface, while
the planes of constant phase are tilted with respect to the same
surface (see Fig. \ref{scheme}). The direction of propagation is
determined by the normal to the constant phase-planes.

Now, for our purpose, it is crucial to note that if the planes of
constant amplitude are parallel to the interface, then
 both the products
$n_2\alpha_2$ and $n_2\beta_2$ in Eq. (\ref{et}) have to be both
real, otherwise the constant-amplitude surfaces would be plane,
but with an inclination with respect to the interface. Moreover,
since the products $n_2\alpha_2$ and $n_2\beta_2$ are real,
$\tan\psi_2$ is also real, since it is the ratio of two real
numbers:

\begin{equation}
\tan\psi_2=\, \frac{n_2\beta_2}{n_2\alpha_2}=
\frac{n_2\sin\theta_2\sin\psi_2}{n_2\sin\theta_2\cos\psi_2}:
\end{equation}
hence, $\psi_2$ is a real angle.

We have therefore obtained a quite unexpected result: even if the
angle of refraction $\theta_2$ in the conducting medium is complex
(as expected) and the refractive index $n_2$ is complex, the
product $n_2\sin\theta_2$ is a real quantity.
 On this basis, by applying the principle of the continuity of the phase
distribution at the interface ($z=0$) between the dielectric and
conducting media, we reobtain the Snell law, that is

\begin{equation}
\label{snell}
  n_2\sin\theta_2=n_0 \sin\theta_0 \:\:\:\: ,
\end{equation}
from which it turns out that $\theta_2$
 is complex:
$\sin\theta_2=\,n_0 \sin\theta_0 /n_2$.
 By solving Eq. (\ref{snell})
we obtain the real $\theta_{2r}$ and imaginary $\theta_{2i}$ parts
of the complex angle of refraction $\theta_{2}$:
\begin{eqnarray}
 &&\sin(\theta_{2r})=
  \left[\frac{ 2B^2n_{2r}^2 } { 1+B^2(n_{2r}^2+n_{2i}^2)+\sqrt{ [1+B^2(n_{2r}^2+n_{2i}^2)]^2-4B^2n_{2r}^2 }
  }\right]^{1/2}
   \nonumber \\
   \nonumber \\
&&  \cosh(\theta_{2i})=
   \left[\frac { 1+B^2(n_{2r}^2+n_{2i}^2)+\sqrt{ [1+B^2(n_{2r}^2+n_{2i}^2)]^2-4B^2n_{2r}^2 }
   }{2}\right]^{1/2}, \nonumber
  \end{eqnarray}
where $n_{2r}$ and $n_{2i}$ are the real and imaginary parts of the
 refractive index $n_2$, respectively, and
 $$B^2=[n_0\sin\theta_0/(n_{2r}^2+n_{2i}^2)]^2.$$

 Coming back to Eq. (\ref{et}), it is expedient to refer to a
cylindrical coordinates system ($\rho,\phi,z$) such that

\begin{equation}
x=\rho\cos\phi,\:\:\:y=\rho\sin\phi, \:\:\: z=z.
\end{equation}
Thus Eqs. (\ref{ei}) and (\ref{et}) can be written as ($n_0=1$)

\begin{eqnarray}
\label{ei1}
&&E_i=\exp[ik_0\rho(\sin\theta_0\cos\psi\cos\phi+\sin\theta_0\sin\psi\sin\phi)]
\exp(ik_0\cos\theta_0z) = \nonumber \\&&=
\exp(ik_0\cos\theta_0z)\,\exp[ik_0\rho\sin\theta_0\cos(\phi-\psi)],
\end{eqnarray}
and

\begin{eqnarray}
\label{et1}
&&E_t=T\exp[ik_0\rho(n_2\sin\theta_2\cos\psi\cos\phi+n_2\sin\theta_2\sin\psi\sin\phi)]\exp(ik_0n_2\cos\theta_2z)
= \nonumber \\
&&=T\exp(ik_0n_2\cos\theta_2z)\,\exp[ik_0\rho\sin\theta_0\cos(\phi-\psi)],
\end{eqnarray}
where use has been made of Eq. (\ref{snell}). The quantity
controlling the $z$-dependence of the wave is complex:

\begin{equation}
n_2\cos\theta_2= \sqrt{n_2^2-\sin^2\theta_0}
\end{equation}
which indicates that the wave attenuates in the direction normal
to the interface ($z$-axis).

Let us now consider a Bessel beam. As is well known, a Bessel beam
originates from the interference of an infinite number of plane
waves whose directions of propagation make the same angle, say
$\theta_0$ , with a given axis, say $z$. Let us suppose to have
such plane waves in medium 1. Therefore, the field propagating in
medium 1 is  given by the sum of an infinite number of waves like
that of Eq. (\ref{ei1}) and results in

\begin{eqnarray}
\label{besi} &&E_i= \exp(ik_0\cos\theta_0z)
\int_0^{2\pi}\exp[ik_0\rho\sin\theta_0\cos(\phi-\psi)]
d\psi=\nonumber \\
 &&=2\pi J_0(k_0\rho\sin\theta_0)\exp(ik_0z\cos\theta_0),
 \end{eqnarray}
where $J_0(k_0\rho\sin\theta_0)$ denotes the zero-order Bessel
function of first kind. Similarly, by integrating Eq. (\ref{et1})
between 0 and $2\pi$, we are able to obtain the field propagating
in medium 2:

\begin{eqnarray}
\label{best} &&E_t= T \exp(ik_0n_2\cos\theta_2z)
\int_0^{2\pi}\exp[ik_0\rho\sin\theta_0\cos(\phi-\psi)]
d\psi=\nonumber \\
 &&=2\pi T J_0(k_0\rho\sin\theta_0)\exp(ik_0n_2\cos\theta_2z)\:.
 \end{eqnarray}
In looking at Eqs. (\ref{besi}) and (\ref{best}), we arrive at
this notable result: a Bessel beam propagating from a perfect
dielectric medium to a conducting one does not change its shape
during the passage. In other words, apart from an attenuation in
the amplitude, no generic Bessel beam - but just the incident beam
- is propagated.

\begin{figure}[h,t]
\begin{center}
\includegraphics[height=10 cm]{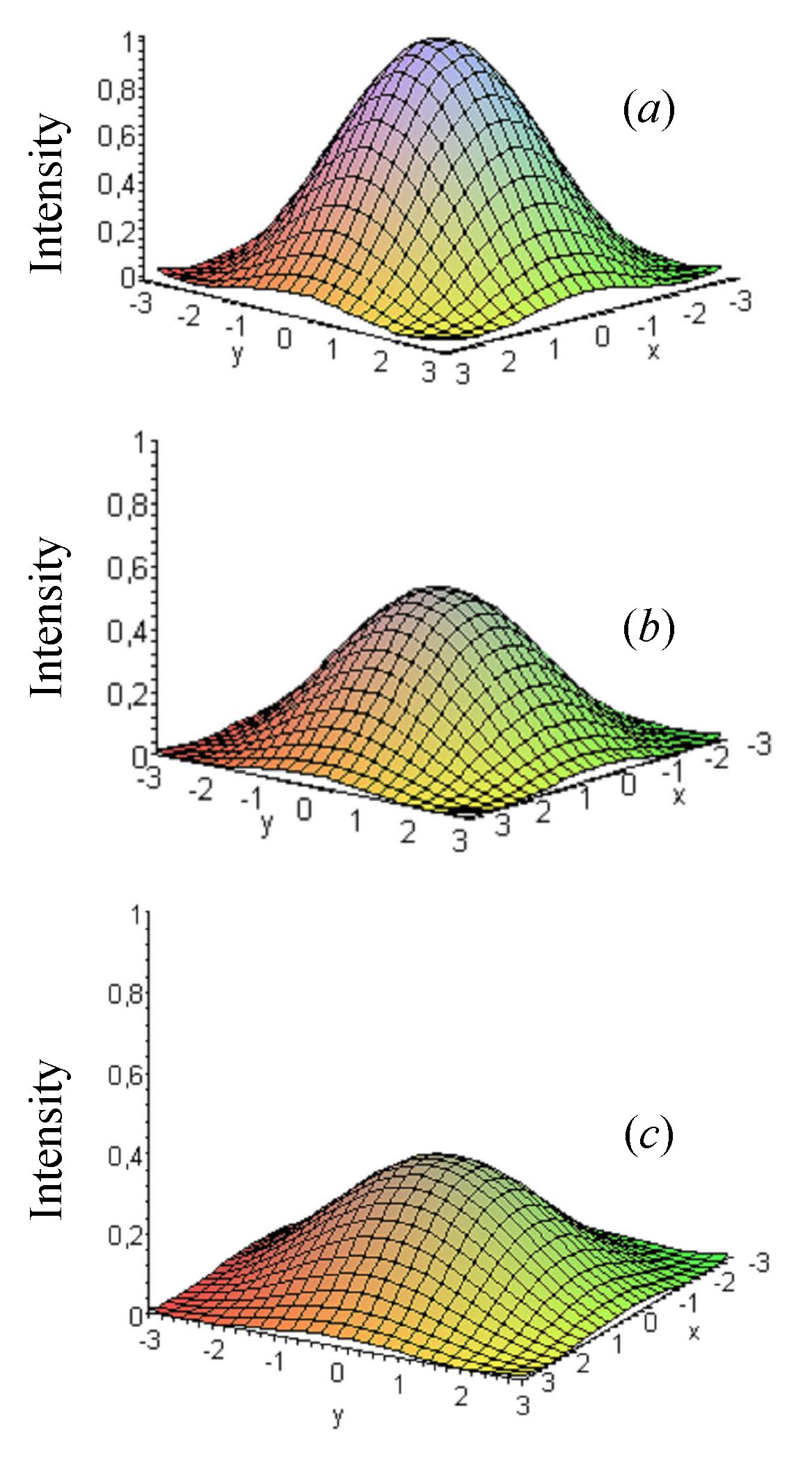}
 \caption{Intensity of the incident Bessel beam at $z=0^-$(\textit{a}) together with the refracted Bessel beam (generated by
refracted inhomogeneous waves) at $z=1$ cm (\textit{b}) and $z=2$ cm
(\textit{c}) inside the conducting medium, as given by Eqs.
(\ref{besi}) and
 (\ref{best}). The transmission coefficient $T =1+(\cos\theta_0-n_2\cos\theta_2)(\cos\theta_0+n_2\cos\theta_2)^{-1}$ has been found by
 applying the continuity conditions of the field and of its first derivative (with respect to the direction of propagation),
 to the border surface between the two media ($z=0$).
 Parameter values are: $\nu=9$ GHz, $\theta_0=20^\circ,\:n_2= 1.2 +i(0.12)$ (a reasonable value for a dry soil).}
 \label{f2}
 \end{center}
 \end{figure}

The field in the conducting medium still has the shape of the
incident Bessel beam: the maxima, the minima and the zeros of the
Bessel function do not change their positions in passing through an
infinite number of conducting layers. The effect of the conducting
media lies only in the attenuation of the amplitude which suffers
some modifications: at $z=0^+$ the amplitude is not 1 but the
complex amplitude $T$, it then attenuates by going away from the
interface (see Fig. \ref{f2}), due to the complex exponential
factor.

When the conductivity increases, the planes of constant phase of
each wave tend to align themselves parallel to the interface. The
effect of the conductivity is to reduce the value of the refraction
angle: for large value of the conductivity the angle of refraction
tends to zero. In this situation the Bessel beam loses its
characteristic of localized wave and tends to become a plane wave.
For infinite conductivity there is no propagation, and the field
inside the conducting medium is zero: the electromagnetic energy is
wholly dissipated by the Joule effect.

\end{document}